\documentclass[10pt,a4paper]{article}
\usepackage[utf8]{inputenc}
\usepackage[english]{babel}

\usepackage{amsmath}
\usepackage{amsfonts}
\usepackage{amssymb}

\usepackage[colorlinks,citecolor=blue,urlcolor=blue,linkcolor=blue]{hyperref}

\usepackage[left=2cm,right=2cm,top=2cm,bottom=2cm]{geometry}

\usepackage{listings}

\usepackage{authblk}

\newcommand{\pd}{\partial}
\newcommand{\uunderline}[1]{\underline{\underline{#1}}}

\title{On simple bootstrap in metric gravity}
\author[1,2]{B. Latosh \thanks{latosh@theor.jinr.ru}}
\affil[1]{Bogoliubov Laboratory of Theoretical Physics, JINR, Dubna 141980, Russia}
\affil[2]{Dubna State University, Universitetskaya str. 19, Dubna 141982, Russia}
\date{}

\begin{document}

\maketitle

\begin{abstract}
  The simplest approach to bootstrap in general relativity is considered. The approach claimed to recover the infinite perturbative series of graviton interaction terms with a recursive coupling of metric perturbations to their energy-momentum tensor. We show that the approach provides an incorrect expression for the term cubic in perturbations. Other difficulties related with the bootstrap approach are discussed.
\end{abstract}

\section{Introduction}\label{introduction}

Explicit calculations in perturbative quantum gravity are obstructed with a major technical difficulty. On the practical level one operates with small metric perturbations $h_{\mu\nu}$ propagating about the flat spacetime $\eta_{\mu\nu}$. The complete spacetime metric $g_{\mu\nu}$ reads
\begin{align}\label{the_perturbative_metric}
  g_{\mu\nu} = \eta_{\mu\nu} + \kappa \, h_{\mu\nu}.
\end{align}
This formula is exact and not a truncation of an infinite series. The field $h_{\mu\nu}$ has the canonical mass dimension and $\kappa$ is a constant with a negative mass dimension. For the sake of convenience we constraint our discussion to general relativity alone and relate $\kappa$ with the Newton constant $G_\text{N}$ as follows:
\begin{align}
  \kappa^2 = 32 \pi\, G_\text{N}.
\end{align}
Perturbations $h_{\mu\nu}$ usually are associated with gravitons, so we sometimes will call them accordingly. Finally, we must also note, that such a setup for perturbation theory is only applicable for an asymptotically flat spaces-time, so the theory has a certain background-dependency.

Within general relativity the dynamics of such perturbation is described by the Hilbert action
\begin{align}\label{Hilbert_action}
  S_H[g_{\mu\nu}] = \int d^4 x \sqrt{-g} \, \left[-\cfrac{1}{16\pi G_\text{N}} \, R\right] 
\end{align}
evaluated on metric \eqref{the_perturbative_metric}. The action produces the following expansion in powers of $h_{\mu\nu}$:
\begin{align}\label{the_action_expansion}
  S_H [ \eta_{\mu\nu} + \kappa\, h_{\mu\nu} ] = S_2 [h_{\mu\nu}] + S_3 [h_{\mu\nu}] + S_4[h_{\mu\nu}] + \cdots \,.
\end{align}
The expansion has no background term as it vanishes on the flat metric. The term linear in perturbations is a complete derivative so it is irrelevant for field equations and can be omitted. Thus the leading expansion terms $S_2$ is quadratic in perturbations and it describes propagation of free perturbations about the flat spacetime. Terms $S_n$ note a part of the action proportional to the $n$-th power of $h_{\mu\nu}$, so $S_n$ term is of an order $\mathcal{O}(\kappa^{n-2})$. In such a way $S_3$ describes three-graviton interaction, $S_4$ describes four-graviton interaction and so on.

The major technical issue behind this approach, for the best of our knowledge, was first highlighted in a series of papers by DeWitt \cite{DeWitt:1967yk,DeWitt:1967ub,DeWitt:1967ucw}. It was noted that the expression for $S_3$ written in a symmetric form contains $171$ terms. For $S_4$ the number of terms rises to $2850$. Such a large number of terms obstructs analytical calculations, so only a handful results were obtained with such analytical calculations \cite{tHooft:1974toh,Goroff:1985th,Grisaru:1975ei,Donoghue:1994dn}.

There is a common believe that it is possible to soften this issue with a certain recursive procedure that recovers the expansion \eqref{the_action_expansion} up to a desired order in perturbation theory. For the best of our knowledge, this believe was founded in the well-known textbook \cite{Misner:1974qy}. For the sake of briefness we call this procedure the ``{\it naive bootstrap}''. The main advantage of the naive bootstrap is the recursive nature, so it is only require to know $S_2$ to recover all $S_n$ up to any given order.

The naive bootstrap operates as follows. Firstly, the energy-momentum tensor $T^{(2)}_{\mu\nu}$ associated with $S_2$ is found. The tensor $T^{(2)}_{\mu\nu}$ is quadratic both in perturbations and derivatives. Secondly, one defines $S_3$ as the energy-momentum tensor $T^{(2)}_{\mu\nu}$ coupled to $h_{\mu\nu}$:
\begin{align}
  S_3 = h^{\mu\nu} \, T^{(2)}_{\mu\nu}.
\end{align}
Next, the energy momentum tensor $T^{(3)}_{\mu\nu}$ associated with $S_3$ is found, and it generates $S_4$ in a similar manner:
\begin{align}
  S_4 = h^{\mu\nu} \, T^{(3)}_{\mu\nu} .
\end{align}
This procedure is repeated until expansion \eqref{the_action_expansion} is recovered up to the desired order. 

In paper \cite{Padmanabhan:2004xk} it was pointed that there are actually no reasons to believe that the naive bootstrap provides the correct expressions for $S_n$.  This sparked a discussion about the existence of any bootstrap mechanism within general relativity and many points raised in \cite{Padmanabhan:2004xk} were addressed in subsequent papers \cite{Butcher:2009ta,Deser:2009fq,Barcelo:2014mua}.

The main aim of this paper is to show that the naive bootstrap described above fails in general relativity. In \cite{Padmanabhan:2004xk} it was argued (among the other) that it is impossible to recover $S_2$ for general relativity by a given $S_1$. We believe that this argument can be improved. One can argue that one cannot obtain the correct expression for $S_2$ from $S_1$ because the linear term is a full derivative, so it does not contribute to field equation. We show that it is not possible to recover $S_3$ from $S_2$ either. We support the claim with explicit calculations.

One of the reason why the naive bootstrap fails is found by a direct comparison with the Yang-Mills theory. Due to the gauge symmetry Yang-Mills theory admits a similar mechanism that allows one to recover three- and four-particle interaction terms by the quadratic part of an action. Within general relativity such a mechanism cannot be easily realized as the quadratic part of an action $S_2$ is invariant with respect to the gauge transformations and requires no nonlinear interaction on the symmetry grounds.

The paper is organized as follows. In Section \ref{naive_bootstrap} we briefly discuss downsides of the naive bootstrap and implement it explicitly to $S_2$. Then we show that the resulting three-graviton action $S_3$ does not match the expression explicitly calculated within perturbation theory. In Section \ref{Yang-Mills_bootstrap} we examine Yang-Mills theory as it enjoys a similar mechanism. Namely, due to the structure of a gauge field variation $\delta A^a_\mu$ (infinitesimal action of the gauge group on a gauge field) it is possible to relate terms with different numbers of fields. Because of this the cubic term is required to cancel a variation of the quadratic part of an action. Quartic term, in term, cancels out the variation of a cubic term. So the complete Yang-Mills Lagrangian is completely defined by its quadratic part. Such a mechanism cannot be realized within simple metric gravity as coordinate transformations (which are an analog of the gauge transformations) act differently on $h_{\mu\nu}$. Unlike the Yang-Mills case the quadratic part of general relativity action has a vanishing variation so no interaction terms should be introduced based on symmetry reasoning. We conclude the paper in Section \ref{discussion} where we address downsides of other possible bootstrap approaches to gravity.

\section{The Naive bootstrap}\label{naive_bootstrap}

Let us implement the naive bootstrap mechanism described in the previous section. First of all, one should define the initial step of the recursive procedure and define $S_2$ which is well-known in the literature \cite{Fierz:1939ix}:
\begin{align}\label{FP}
  S_{(2)}=\int d^4 x \left[\, \cfrac12\,\pd_\lambda h_{\rho\sigma} \,\pd^\lambda h^{\rho\sigma} - \cfrac12\, \pd_\lambda h \,\pd^\lambda h + \pd_\rho h^{\rho\sigma} \, \pd_\sigma h - \pd_\lambda h_{\rho\sigma} \,\pd^\rho h^{\sigma\lambda} \right] .
\end{align}
The action is fixed uniquely. Firstly, it can be recovered as the most general action for rank-$2$ symmetric tensor field \cite{Fierz:1939ix}. Secondly, it can be directly recovered from the perturbative general relativity expansion.
%

The second step is to evaluate the energy-momentum tensor associated with $S_2$. The canonical expression for an energy-momentum tensor of an arbitrary field $\phi$ with a Lagrangian $\mathcal{L}$ reads
\begin{align}\label{canonical_EMT}
  T_{\mu\nu} = \cfrac{\delta \mathcal{L}}{\delta (\pd^\mu \phi)}\, \pd_\nu \phi - \eta_{\mu\nu} \, \mathcal{L} .
\end{align}
If field equations are satisfied, then the tensor is conserved:
\begin{align}
  \pd^\mu T_{\mu\nu} = \cfrac{\delta\mathcal{L}}{\delta \phi}\, \pd_\nu \phi + \cfrac{\delta\mathcal{L}}{\delta (\pd^\mu \phi)} \, \pd_\nu (\pd^\mu\phi) - \pd_\nu \mathcal{L} =0 .
\end{align}
The first two terms constitute the full derivative $\pd_\nu \mathcal{L}$ which is cancel out by the last term. We use \eqref{canonical_EMT} as the expression defining the energy-momentum tensor.

The main issue with such a definition of an energy-momentum tensor is the fact that it is not unique. If one takes an arbitrary tensor $t_{\alpha\mu\nu}$ which is anti-symmetric with respect to the first two indices $t_{\alpha\mu\nu} = - t_{\mu\alpha\nu}$, then $T_{\mu\nu} + \pd^\alpha t_{\alpha\mu\nu}$ is also conserved. Therefore one ends up with an infinite number of conserved tensors generated by the expression \eqref{canonical_EMT}. The non-uniqueness of the energy-momentum tensor does not pose a problem on its own.
The technique of superpotentials allows one to implement such a non-unique definition of energy-momentum tensor for practical calculations \cite{Giachetta:1995bj,Petrov:2002mra,Petrov:2020abu,Petrov:2012qn}. But the non-uniqueness of an energy-momentum tensor is crucial for the naive bootstrap as the tensor is coupled to small perturbations $h_{\mu\nu}$. A redefinition of the energy-momentum tensor
\begin{align}
  T_{\mu\nu} \to T_{\mu\nu} + \pd^\alpha t_{\alpha\mu\nu} \text{ with } t_{\alpha\mu\nu} = - t_{\mu\alpha\nu}
\end{align}
may be expected to be irrelevant for bootstrapped $S_3$ if and only if the contribution associated with $t_{\alpha\mu\nu}$ is reduced to surface terms:
\begin{align}
  \int d^4 x \, h^{\mu\nu} \, \pd^\alpha\, t_{\alpha\mu\nu} \to \text{ surface terms}.
\end{align}
For an arbitrary $t_{\alpha\mu\nu}$ this is not the case.

Because of this the naive bootstrap in the given form is not complete. It must be completed with a condition defining the correct expression for the energy-momentum tensor that provides the correct expression for $S_3$. The absence of this condition alone makes the naive bootstrap a poorly defined procedure. One may argue that gauge-invariance of the energy-momentum tensor is the correct condition to proceed, but this is not the case, as there is no gauge-invariant definition of the energy-momentum tensor for linear gravity \cite{Magnano:2002xx}. It also must be highlighted that there are numerous issues related with the energy-momentum tensor within linear gravity, so the issue of a proper choice of energy-momentum tensor is far from being resolved \cite{Baker:2021qqi,Baker:2020eqs}.

To proceed with an implementation of the naive bootstrap we have no choice, but to resolve this issue. It is natural to use the canonical definition of the energy-momentum tensor \eqref{canonical_EMT} as it is provided by the Noether's theorem \cite{Noether:1918zz}. 
The next step is to evaluate the energy-momentum tensor \eqref{canonical_EMT} for $S_2$ \eqref{FP} explicitly:
\begin{align}\label{T2_bootstrap}
  \begin{split}
    T^{(2)}_{\mu\nu}  &= \cfrac{\delta\mathcal{L}}{\delta (\pd^\mu h_{\alpha\beta})} \, \pd_\nu h_{\alpha\beta} - \eta_{\mu\nu} \mathcal{L}_2\\
    &= \pd_\mu h^{\alpha\beta}\, \pd_\nu h_{\alpha\beta} - \pd_\mu h\, \pd_\nu h \, +\pd_\nu h_{\mu\sigma}\, \pd^\sigma h + \pd_\nu h \, \pd^\sigma h_{\sigma\mu} -2 \pd_\nu h_{\alpha\beta}\, \pd^\alpha \, h^\beta{}_\mu - \eta_{\mu\nu} \mathcal{L}_2 \,.
  \end{split}
\end{align}
This expression is not symmetric with respect to $\mu \leftrightarrow \nu$, but when coupled to $h_{\mu\nu}$ only the symmetric part remains in the action. At the same time, both symmetric and anti-symmetric parts are required for the conservation of $\pd^\mu T^{(2)}_{\mu\nu}=0$.
This appears not to post any problem what so ever. Action $S_3$ describes three-graviton interaction, so it is natural to expect energy-momentum transfer between interacting perturbations.

Finally, $T^{(2)}_{\mu\nu}$ should be coupled to $h_{\mu\nu}$ to form $S_3$. In order to distinguish the action obtained via the naive bootstrap we note it as $S_3^\text{NB}$:
\begin{align}\label{S3_NB}
  \begin{split}
    S_3^\text{NB} =& h^{\mu\nu} \, T^{(2)}_{\mu\nu} \\
    =& h^{\mu\nu}\,\pd_\mu h_{\alpha\beta}\, \pd_\nu h^{\alpha\beta} - h^{\mu\nu} \, \pd_\mu h \, \pd_\nu h + h^{\mu\nu} \, \pd_\mu h_{\nu\sigma} \, \pd^\sigma h + h^{\mu\nu} \,\pd_\mu h \, \pd^\rho h_{\rho\nu} \\
    & -2 h^{\mu\nu} \, \pd_\mu h^{\alpha\beta} \, \pd_\alpha h_{\nu\beta}-\cfrac12\, h \, \pd_\mu h_{\alpha\beta} \, \pd^\mu h^{\alpha\beta} + \cfrac12\, h \, \pd_\mu h \, \pd^\mu h - h \, \pd_\rho h^{\rho\sigma} \, \pd_\sigma h   + h \,\pd_\mu h_{\nu\sigma} \, \pd^\nu h^{\mu\sigma}.
  \end{split}
\end{align}

Expression \eqref{S3_NB} can be compared directly with the expression obtained with perturbative calculations. This is enough to show that the naive bootstrap fails to provide the correct answer. Firstly, one should pay attention to the dimension of expression \eqref{S3_NB}. We used the natural units $c=\hbar =1$ throughout the paper, so $S_2$ has the correct mass dimension $4$. The given expression for $S_3^\text{NB}$ \eqref{S3_NB} has mass dimension $5$, thus it must be divided by a suitable mass scale. The naive bootstrap provides no reasons why this should be the Planck scale. At the level of three-graviton interaction this problem is avoided as the required mass scale $M_3$ can always be labeled as ``the Planck scale''. Therefore one can always claim that the following expression has the correct mass dimension and new mass scale $M_3$ is the Planck mass:
\begin{align}
  \cfrac{1}{M_3} \, S_3^\text{NB}.
\end{align}
But the problem reemerges at the level of $S_4$. The energy-momentum tensor $T^{(3)}$ evaluated with $M_3^{-1}\,S_3^\text{NB}$ also have the mass dimension equal to $4$. Thus $S_4^\text{NB} = T^{(3)}_{\mu\nu} \, h^{\mu\nu}$ has the mass dimension $5$ and it must be divided on a new independent mass scale $M_4$. There are no reasons to believe that $M_3$ and $M_4$ are the same scales. Moreover, there are no reasons to believe that they are of the same order either. Because of this within the simple bootstrap approach each $S_n$ has its own independent mass scale. Unless some fine tuning mechanism is introduced, there are simply no reasons to believe that these scales are of the same order of magnitude.

The most important issue of the naive bootstrap is founded when expression expression \eqref{S3_NB} is directly compared with the one obtained with perturbative calculations:
\begin{align}\label{S3}
  \begin{split}
    \left(\int d^4 x \sqrt{-g} R\right)^{(3)} =&\int d^4 x \left(\cfrac{\kappa^3}{4}\right) \Bigg[  h^{\mu\nu} \,\pd_\mu h_{\rho\sigma} \,\pd_\nu h^{\rho\sigma} - 2\, h^{\mu\nu}\,\pd_\rho h_{\mu\sigma}\, \pd^\sigma h_\nu{}^\rho+2 \, h^{\mu\nu} \, \pd_\rho h_{\mu\sigma}\,\pd^\rho h_\nu {}^\sigma \\
      & - 4\, h^{\mu\nu}\, \pd_\mu h_{\rho\sigma} \, \pd^\sigma h_\nu{}^\rho +2 \, h^{\mu\nu} \, \pd_\mu h_{\nu\sigma} \, \pd^\sigma h - 2 \, h^{\mu\nu} \, \pd_\sigma h_{\mu\nu} \, \pd^\sigma h \\
      &+ 2 \,  h^{\mu\nu}\,\pd_\sigma h_{\mu\nu} \, \pd_\rho h^{\rho\sigma} + 2 \, h^{\mu\nu}\,\pd_\mu h \,\pd^\sigma h_{\sigma\nu}- h^{\mu\nu}\,\pd_\mu h\, \pd_\nu h\\
      &-\cfrac12 \, h\, \pd_\lambda h_{\mu\nu}\,\pd^\lambda h^{\mu\nu} +\cfrac12 \, h\,\pd_\sigma h\, \pd^\sigma h  - h\, \pd_\sigma h \,\pd_\rho h^{\rho\sigma} + h\, \pd_\mu h_{\nu\sigma}\,\pd^\nu h^{\mu\sigma} \Bigg].
  \end{split}
\end{align}
This expression was derived manually, as described below, and independently checked with a system of computer algebra. It must be noted that a similar expression presents in paper \cite{Goroff:1985th}. However, it (equation (A.5) of \cite{Goroff:1985th}) has a misprint in the coefficient of $h \pd_\mu \pd^\mu h$ term.

Expression \eqref{S3} is obtained as follows. Firstly, surface terms should be separated:
\begin{align}
  \int d^4 x\sqrt{-g} R = \int d^4 x  \sqrt{-g} g^{\mu\nu} \left[ \Gamma_{\mu\sigma}^\rho \, \Gamma_{\nu\rho}^\sigma - \Gamma_{\mu\nu}^\sigma \,\Gamma_{\sigma\rho}^\rho \right] + \text{surface terms} .
\end{align}
We use notations from \cite{tHooft:1974toh} and mark background quantities with a single line above a symbol; quantities linear in $h_{\mu\nu}$ we mark with a single line below a symbol; quantities quadratic in perturbations we mark with two lines below a symbol, etc. This allows one to write the expression as follows:
\begin{align}
  \begin{split}
    \int d^4 x \underline{\underline{\underline{\sqrt{-g} \, R}}} =\int d^4 x \Bigg[& \underline{\sqrt{-g} \, g^{\mu\nu}} \left[ \underline{\Gamma^\sigma_{\mu\rho}}\,\underline{\Gamma^\rho_{\nu\sigma}} - \underline{\Gamma^\sigma_{\mu\nu}} \, \underline{\Gamma^\rho_{\sigma\rho}} \right] \\
      &+ \sqrt{-\overline{g}} \, \overline{g}^{\mu\nu} \left[ \uunderline{\Gamma^\sigma_{\mu\rho}}\,\underline{\Gamma^\rho_{\nu\sigma}} - \uunderline{\Gamma^\sigma_{\mu\nu}} \, \underline{\Gamma^\rho_{\sigma\rho}} + \underline{\Gamma^\sigma_{\mu\rho}}\,\uunderline{\Gamma^\rho_{\nu\sigma}} - \underline{\Gamma^\sigma_{\mu\nu}} \, \uunderline{\Gamma^\rho_{\sigma\rho}} \right] \Bigg].
  \end{split}
\end{align}
This expression it is much simpler computational wise because $\overline{\Gamma}^\alpha_{\mu\nu}$ vanishes. Consequently, one is actually relived from a calculation of terms cubic in $\kappa$. Finally, the following formulae for perturbative expansions should be used:
\begin{align}
  \begin{split}
    \sqrt{-g} g^{\mu\nu} &= \eta^{\mu\nu} -\kappa \left(h^{\mu\nu} - \cfrac12\, h \,\eta^{\mu\nu} \right) + \mathcal{O}(\kappa^2), \\
    \Gamma_{\mu\nu}^\alpha &= \cfrac{\kappa}{2}\left[ \pd_\mu h_\nu{}^\alpha+\pd_\nu h_\mu{}^\alpha -\pd^\alpha h_{\mu\nu}\right] - \cfrac{\kappa^2}{2}\,h^{\alpha\beta}\left[\pd_\mu h_{\nu\beta}+\pd_\nu h_{\mu\beta} - \pd_\beta h_{\mu\nu} \right] + \mathcal{O}(\kappa^3).
  \end{split}
\end{align}
These expressions combined result in formula \eqref{S3}.

Expressions \eqref{S3} and \eqref{S3_NB} shall be compared directly as follows:
\begin{align}\label{the_difference}
  \begin{split}
    \left(\cfrac{4}{\kappa^3} \int d^4 x \sqrt{-g} \,R\right) - S_3^\text{NB} =& 2 h^{\mu\nu} \left[ - \pd_\mu h_{\rho\sigma}\, \pd^\rho h_\nu{}^\sigma+ \pd_\sigma h_{\mu\nu} \, \pd_\rho h^{\rho\sigma} - \pd_\rho h_{\mu\sigma}\, \pd^\sigma h_\nu{}^\rho+ \pd_\rho h_{\mu\sigma}\,\pd^\rho h_\nu {}^\sigma\right] \\
    &+ \pd^\mu h \left[h^{\rho\sigma} \pd_\rho h_{\sigma\mu} + h_{\mu\sigma} \pd_\rho h^{\rho\sigma} -2 h^{\rho\sigma} \pd_\mu h_{\rho\sigma}\right].
  \end{split}
\end{align}
This expression helps one to see that the difference between \eqref{S3} and \eqref{S3_NB} is not a full derivative. One may expect that this expression either vanish or reduces to a full derivative at least when linear field equations are satisfied. However, this is also not the case. The difference neither vanish nor reduces to a complete derivative even if we require $h=0$, $\pd_\mu h^{\mu\nu}=0$, $\square h_{\mu\nu}=0$. Which support our original claim that it is impossible to obtain the correct expression for $S_3$ from $S_2$.

Let us summarize all downsides of the naive bootstrap approach found in this paper. Firstly, the naive bootstrap lacks a method to define mass scales of each interaction term $S_n$. Secondly, in the current form the approach lacks a well-stated definition of the energy-momentum tensor and, as it was highlighted, the issue goes far beyond the naive bootstrap. Finally, if one uses the canonical energy-momentum tensor \eqref{canonical_EMT} to obtain the explicit expression for $S_3$, then the obtained result \eqref{S3_NB} does not match the the correct expression \eqref{S3} obtained from the perturbative expansion.

In the next section we explain why the naive bootstrap fails on an example of Yang-Mills theory where it is actually possible to recover the nonlinear part of a Lagrangian by the kinetic term.

\section{Yang-Mills bootstrap}\label{Yang-Mills_bootstrap}

We believe that the main reason why the simple bootstrap fails is uncovered via a careful comparison with the Yang-Mills theory. The theory enjoys a similar mechanism that allows one to recover three- and four-particle Lagrangians from the kinetic term. The mechanism is due to the gauge symmetry and the structure of a group action on a gauge field.

Let us define Yang-Mills field tensor as follows:
\begin{align}
  F^c_{\mu\nu} = \pd_\mu A^c_\nu -\pd_\nu A^c_\mu + C_{abc} \, A^a_\mu\, A^b_\nu .
\end{align}
Here $A^a_\mu$ is a gauge field with a Lorentz index $\mu$ and a group index $a$. We note the group generators as $T^a$ and define the structure constants $C_{abc}$ as follows:
\begin{align}
  [T_a, T_b ] = C_{abc} \, T_c .
\end{align}

In such a parametrization the infinitesimal action of a gauge group on a gauge field (variation of a gauge field) reads
\begin{align}\label{YM_IA_DEF}
  \delta A^c_\mu = i \omega^a \, A^b_\mu \, C_{abc} - i \pd_\mu \omega^c ,
\end{align}
with $\omega^a = \omega^a(x)$ begin the gauge parameter. In this expression the first term does contain the gauge field $A^a_\mu$ while the second one does not. Because of this the variation of the part of a Lagrangian containing $n$ fields produce terms which contain both $n$ and $n-1$ fields:
\begin{align}
  \delta (A^n) \to \omega \, A^n + \pd\omega \, A^{n-1}.
\end{align}
Consequently, terms with different powers of the gauge field are related and the full nonlinear Lagrangian is fixed.

Let us show this explicitly. The Yang-Mills Lagrangian reads
\begin{align}
  \mathcal{L} = -\cfrac14 \, F^c_{\mu\nu} \, F^c {}^{\mu\nu} = -\cfrac14\, f^c_{\mu\nu} \, f^c{}^{\mu\nu} - \cfrac12 \, f^c{}^{\mu\nu} \, C_{abc} \, A^a_\mu \, A^b_\nu -\cfrac14\, C_{abc} \,C_{nmc} \, A^{a\mu}\, A^{b\,\nu}\,A^m_\mu \, A^n_\nu  ,
\end{align}
with $f^c_{\mu\nu}$ defined as follows:
\begin{align}
  f^c_{\mu\nu} \overset{\text{def}}{=} \pd_\mu A^c_\nu - \pd_\nu A^c_\mu.
\end{align}
The infinitesimal group action on quadratic, cubic, and quartic parts of the Lagrangian reads:
\begin{align}\label{YM_IA_2}
  \delta\left( -\cfrac14 \, f^c_{\mu\nu} \, f^c{}^{\mu\nu} \right) =& i \omega^a\, C_{abc} \, A^b_\nu \,\pd_\mu f^c{}^{\mu\nu} , \hspace{9.4cm}
\end{align}
\begin{align}\label{YM_IA_3}
  \delta\left(- \cfrac12 \, f^c{}^{\mu\nu} \, C_{abc} \, A^a_\mu \, A^b_\nu\right) =& -i \omega^a\, C_{abc} \, A^b_\nu \,\pd_\mu f^c{}^{\mu\nu}   + i \omega^r \, C_{rsc}\, A^{s\nu} \pd^\mu \left( C_{abc}\,A^a_\mu \, A^b_\nu\right)-\cfrac{i}{2}\, \omega^a \, C_{abc} \, F^b_{\mu\nu} \, f^{c\,\mu\nu} ,
\end{align}
\begin{align}\label{YM_IA_4}
  \delta\left(-\cfrac14\, C_{abc} \,C_{nmc} \, A^{a\mu}\, A^{b\,\nu}\,A^m_\mu \, A^n_\nu  \right) =& -i \omega^r \, C_{rsc}\, A^{s\nu} \pd^\mu \left(C_{abc}\,A^a_\mu \, A^b_\nu\right) -\cfrac{i}{2}\, \omega^a \, C_{abc} F^b_{\mu\nu}\left( C_{mnc} \, A^{m\,\mu} \, A^{n\,\nu}\right) ,
\end{align}
The quadratic part has a non-vanishing variation \eqref{YM_IA_2}. Consequently, it cannot enter an action alone as it must be invariant with respect to the gauge group. To cancel out \eqref{YM_IA_2} one must introduce a three-particle interaction and its form is fixed uniquely as it must perform the cancellation. However, the sum of these terms also has a non-vanishing variation, so a quartic Lagrangian shall be introduced and its form, once again, is fixed uniquely. Only the sum of quadratic, cubic, and quartic terms enjoy a vanishing variation due to the symmetry of the structure constants:
\begin{align}
  \delta \mathcal{L} = -\cfrac{i}{2} \, \omega^a \, C_{abc} \, F^b_{\mu\nu} \, F^c{}^{\mu\nu} =0 .  
\end{align}

The core of this mechanism is the form of an infinitesimal group action \eqref{YM_IA_DEF}. It is possible to recover the complete Lagrangian solely because \eqref{YM_IA_DEF} relates terms with different powers of a gauge field. The structure of infinitesimal group action for gravity has no room for such a mechanism. For gravity coordinate transformations $x^\mu \to x^\mu + \zeta^\mu$ play the role of gauge transformations. Their infinitesimal action on small metric perturbations $h_{\mu\nu}$ does not contains the perturbations themselves:
\begin{align}
  \delta h_{\mu\nu} = \pd_\mu \zeta_\nu + \pd_\nu \zeta_\mu .
\end{align}
Therefore an infinitesimal element of a term containing $n$ metric perturbations will always produce a contribution with only $n-1$ perturbations:
\begin{align}
  \delta ( h^n) \to \zeta \, \pd(h^{n-1}).
\end{align}
This fact by itself renders out an opportunity to find a simple bootstrap mechanism similar to the Yang-Mills case as one simple does not require terms with higher powers of $h_{\mu\nu}$ to end up with an invariant Lagrangian. Finally, the quadratic Lagrangian $S_2$ which is obtained from the action perturbative expansion admits a vanishing infinitesimal group action
\begin{align}
  \delta\mathcal{L}_\text{FP} =0 .
\end{align} 

This shows that gravity has no room for a simple bootstrap mechanism. Firstly, the quadratic part of an action is invariant and requires no interaction terms on the symmetry grounds. Secondly, even if an $n$-graviton interaction terms is introduced, it may only require terms with a lesser power of $h_{\mu\nu}$. Therefore, in such a framework the symmetry provides no way to generate an infinite number of interaction terms.

\section{Discussion}\label{discussion}

We studied a procedure called naive bootstrap, described in Section \ref{introduction}, that supposed to provide a simple way to recover the perturbative expansion of the Hilbert action. We expanded one argument presented in \cite{Padmanabhan:2004xk}. It was claimed that the naive bootstrap fails to recover the correct quadratic part of the gravity action by a given linear part. We believe that this argument may be negated as the liner part of an action is a complete derivative. We evaluated the cubic part of the Lagrangian with the naive bootstrap \eqref{S3_NB} and it does not match the correct expression \eqref{S3} obtained with perturbative calculations. To be exact, the difference between these expressions \eqref{the_difference} neither vanishes, nor becomes a complete derivative even if we assume $h=0$, $\pd_\mu h^{\mu\nu}=0$, $\square h_{\mu\nu}=0$. This explicitly shows that the naive bootstrap fails.

We identified a series of disadvantages of the naive bootstrap that, as we believe, is the reason behind its inapplicability. First and foremost, there are no theoretical reasons to introduce interaction to the linear theory given by action \eqref{FP}. The action \eqref{FP}, which is obtained from perturbative expansion of general relativity, is invariant with respect to gauge transformation. Therefore symmetry does not provide a way to relate $S_2$ and $S_3$, so one can hardly expect to find any mechanism in spirit of the naive bootstrap. Secondly, in the current form the naive bootstrap lacks a suitable definition of the energy-momentum tensor which supposed to be coupled to $h_{\mu\nu}$ to form $S_3$. Moreover, as we noted in Section \ref{naive_bootstrap}, the naive bootstrap lacks a way to generate energy scales for each $S_n$. Strictly speaking, this makes the whole naive bootstrap being poorly defined. Finally, even if we prefer to ignore all its downsides and implement the naive bootstrap for $S_3$, then we obtain \eqref{S3_NB} which cannot be reduced to the correct expression \eqref{S3}.

Thus we conclude that the simplest approach to gravity bootstrap, discussed in Section \ref{introduction}, fails. However, it shall be noted that the whole bootstrap paradigm must not be reduced to mechanism is spirit on the naive bootstrap. Namely, it is established that the cubic interaction described by general relativity \eqref{S3} is the only interaction consistent with causality \cite{Camanho:2014apa,Meltzer:2017rtf,Belin:2019mnx,Kologlu:2019bco}. And the spin principle also provides a way to reconstruct the complete nonlinear gravity theory with a certain assumption on the structure of its interaction \cite{Ogievetsky:1964pss,Ogievetsky:1965zcd}.

\section*{Acknowledgment}
  The work was supported by the Foundation for the Advancement of Theoretical Physics and Mathematics “BASIS”.

\bibliographystyle{unsrt}
\bibliography{Bootstrap.bib}

\begin{thebibliography}{10}

\bibitem{DeWitt:1967yk}
Bryce~S. DeWitt.
\newblock {Quantum Theory of Gravity. 1. The Canonical Theory}.
\newblock {\em Phys. Rev.}, 160:1113--1148, 1967.
\newblock [3,93(1987)].

\bibitem{DeWitt:1967ub}
Bryce~S. DeWitt.
\newblock {Quantum Theory of Gravity. 2. The Manifestly Covariant Theory}.
\newblock {\em Phys. Rev.}, 162:1195--1239, 1967.
\newblock [,298(1967)].

\bibitem{DeWitt:1967ucw}
Bryce~S. DeWitt.
\newblock {Quantum Theory of Gravity. 3. Applications of the Covariant Theory}.
\newblock {\em Phys. Rev.}, 162:1239--1256, 1967.
\newblock [,307(1967)].

\bibitem{tHooft:1974toh}
Gerard 't~Hooft and M.~J.~G. Veltman.
\newblock {One loop divergencies in the theory of gravitation}.
\newblock {\em Ann. Inst. H. Poincare Phys. Theor.}, A20:69--94, 1974.

\bibitem{Goroff:1985th}
Marc~H. Goroff and Augusto Sagnotti.
\newblock {The Ultraviolet Behavior of Einstein Gravity}.
\newblock {\em Nucl. Phys.}, B266:709--736, 1986.

\bibitem{Grisaru:1975ei}
Marcus~T. Grisaru, P.~van Nieuwenhuizen, and C.C. Wu.
\newblock {Background Field Method Versus Normal Field Theory in Explicit
  Examples: One Loop Divergences in S Matrix and Green's Functions for
  Yang-Mills and Gravitational Fields}.
\newblock {\em Phys. Rev. D}, 12:3203, 1975.

\bibitem{Donoghue:1994dn}
John~F. Donoghue.
\newblock {General relativity as an effective field theory: The leading quantum
  corrections}.
\newblock {\em Phys. Rev.}, D50:3874--3888, 1994.

\bibitem{Misner:1974qy}
Charles~W. Misner, K.~S. Thorne, and J.~A. Wheeler.
\newblock {\em {Gravitation}}.
\newblock W. H. Freeman, San Francisco, 1973.

\bibitem{Padmanabhan:2004xk}
T.~Padmanabhan.
\newblock {From gravitons to gravity: Myths and reality}.
\newblock {\em Int. J. Mod. Phys. D}, 17:367--398, 2008.

\bibitem{Butcher:2009ta}
Luke~M. Butcher, Michael Hobson, and Anthony Lasenby.
\newblock {Bootstrapping gravity: A Consistent approach to energy-momentum
  self-coupling}.
\newblock {\em Phys. Rev. D}, 80:084014, 2009.

\bibitem{Deser:2009fq}
S.~Deser.
\newblock {Gravity from self-interaction redux}.
\newblock {\em Gen. Rel. Grav.}, 42:641--646, 2010.

\bibitem{Barcelo:2014mua}
Carlos Barcel\'o, Ra\'ul Carballo-Rubio, and Luis~J. Garay.
\newblock {Unimodular gravity and general relativity from graviton
  self-interactions}.
\newblock {\em Phys. Rev. D}, 89(12):124019, 2014.

\bibitem{Fierz:1939ix}
M.~Fierz and W.~Pauli.
\newblock {On relativistic wave equations for particles of arbitrary spin in an
  electromagnetic field}.
\newblock {\em Proc. Roy. Soc. Lond. A}, 173:211--232, 1939.

\bibitem{Giachetta:1995bj}
Giovanni Giachetta and Gennadi Sardanashvily.
\newblock {Stress energy momentum of affine metric gravity. Generalized Komar
  superpotential}.
\newblock {\em Class. Quant. Grav.}, 13:L67--L72, 1996.

\bibitem{Petrov:2002mra}
Alexander~N. Petrov and Joseph Katz.
\newblock {Conserved currents, superpotentials and cosmological perturbations}.
\newblock {\em Proc. Roy. Soc. Lond. A}, 458(2018):319--337, 2002.

\bibitem{Petrov:2020abu}
A.~N. Petrov and J.~Brian Pitts.
\newblock {The Field-Theoretic Approach in General Relativity and Other Metric
  Theories. A Review}.
\newblock 4 2020.

\bibitem{Petrov:2012qn}
Alexander~N. Petrov and Robert~R. Lompay.
\newblock {Covariantized Noether identities and conservation laws for
  perturbations in metric theories of gravity}.
\newblock {\em Gen. Rel. Grav.}, 45:545--579, 2013.

\bibitem{Magnano:2002xx}
Guido Magnano and Leszek~M. Sokolowski.
\newblock {Symmetry properties under arbitrary field redefinitions of the
  metric energy-momentum tensor in classical field theories and gravity}.
\newblock {\em Class. Quant. Grav.}, 19:223--236, 2002.

\bibitem{Baker:2021qqi}
Mark~Robert Baker.
\newblock {Canonical Noether and the energy-momentum non-uniqueness problem in
  linearized gravity}.
\newblock {\em Class. Quant. Grav.}, 38(9):095007, 2021.

\bibitem{Baker:2020eqs}
Mark~Robert Baker, Natalia Kiriushcheva, and Sergei Kuzmin.
\newblock {Noether and Hilbert (metric) energy-momentum tensors are not, in
  general, equivalent}.
\newblock {\em Nucl. Phys. B}, 962:115240, 2021.

\bibitem{Noether:1918zz}
Emmy Noether.
\newblock {Invariant Variation Problems}.
\newblock {\em Gott. Nachr.}, 1918:235--257, 1918.

\bibitem{Camanho:2014apa}
Xian~O. Camanho, Jose~D. Edelstein, Juan Maldacena, and Alexander Zhiboedov.
\newblock {Causality Constraints on Corrections to the Graviton Three-Point
  Coupling}.
\newblock {\em JHEP}, 02:020, 2016.

\bibitem{Meltzer:2017rtf}
David Meltzer and Eric Perlmutter.
\newblock {Beyond $a = c$: gravitational couplings to matter and the stress
  tensor OPE}.
\newblock {\em JHEP}, 07:157, 2018.

\bibitem{Belin:2019mnx}
Alexandre Belin, Diego~M. Hofman, and Gr\'egoire Mathys.
\newblock {Einstein gravity from ANEC correlators}.
\newblock {\em JHEP}, 08:032, 2019.

\bibitem{Kologlu:2019bco}
Murat Kologlu, Petr Kravchuk, David Simmons-Duffin, and Alexander Zhiboedov.
\newblock {Shocks, Superconvergence, and a Stringy Equivalence Principle}.
\newblock {\em JHEP}, 11:096, 2020.

\bibitem{Ogievetsky:1964pss}
V.~I. Ogievetsky and I.~V. Polubarinov.
\newblock {On interacting fields with definite spin}.
\newblock {\em Zh. Eksp. Teor. Fiz.}, 45(1):237--245, 1963.

\bibitem{Ogievetsky:1965zcd}
V.~I Ogievetsky and I.~V Polubarinov.
\newblock {Interacting field of spin 2 and the Einstein equations}.
\newblock {\em Annals Phys.}, 35(2):167--208, 1965.

\end{thebibliography}

\end{document}